\documentclass[twocolumn,showpacs,aps,prl,superscriptaddress]{revtex4}
\usepackage{graphicx}
\usepackage{dcolumn}
\usepackage{amsmath}
\usepackage{epsfig}
\include{babarsym}

\def\figurebox#1#2#3{
    \def\arg{#3}
    \ifx\arg\empty
    {\hfill\vbox{\hsize#2\hrule\hbox to #2{\vrule\hfill\vbox to #1{\hsize#2\vfill}\vrule}\hrule}\hfill}
    \else
    {\hfill\epsfbox{#3}\hfill}
    \fi}

\begin{document}

\noindent\babar-PUB-08/030\\
SLAC-PUB-13375\\

\title{\large \bf Evidence for \boldmath$X(3872)\rightarrow\psitwos\gamma$ in \boldmath$\Bpm\rightarrow X(3872)\Kpm$ decays,\\and a study of $\B\rightarrow\ccbar\gamma K$}

%
\author{B.~Aubert}
\author{M.~Bona}
\author{Y.~Karyotakis}
\author{J.~P.~Lees}
\author{V.~Poireau}
\author{E.~Prencipe}
\author{X.~Prudent}
\author{V.~Tisserand}
\affiliation{Laboratoire de Physique des Particules, IN2P3/CNRS et Universit\'e de Savoie, F-74941 Annecy-Le-Vieux, France }
\author{J.~Garra~Tico}
\author{E.~Grauges}
\affiliation{Universitat de Barcelona, Facultat de Fisica, Departament ECM, E-08028 Barcelona, Spain }
% ...
\collaboration{The \babar\ Collaboration}
\noaffiliation

\begin{abstract}
\noindent In a search for $B\rightarrow \ccbar\gamma K$ decays with the \babar\ detector, where $\ccbar$ includes $J/\psi$ and $\psi(2S)$, and $K$ includes \Kpm, \KS and $\Kstar (892)$, we find evidence for $X(3872)\rightarrow J/\psi\gamma$ and $X(3872)\rightarrow \psi(2S)\gamma$ with $3.6\sigma$ and $3.5\sigma$ significance, respectively. We measure the product of branching fractions $\BR(\Bpm\rightarrow X(3872)\Kpm)\cdot\BR(X(3872)\rightarrow J/\psi\gamma) = (2.8\pm0.8(stat.)\pm0.1(syst.))\times10^{-6}$ and $\BR(\Bpm\rightarrow X(3872)\Kpm)\cdot\BR(X(3872)\rightarrow \psitwos\gamma) = (9.5\pm2.7(stat.)\pm0.6(syst.))\times10^{-6}$.
\end{abstract}


\maketitle

\def\bfDbar    {\kern 0.2em\overline{\kern -0.2em D}{}\xspace}
\def\bfDzb     {\ensuremath{\bfDbar^0}\xspace}
\def\bfDstarz  {\ensuremath{D^{*0}}\xspace}
\def\bfDzDzb   {\ensuremath{\bfDzb {\kern -0.16em \bfDstarz}}\xspace}

\noindent The $X(3872)$ state discovered by the Belle Collaboration in the decay $\Bpm\rightarrow\Kpm X(3872), X(3872)\rightarrow\jpsi\pi^{+}\pi^{-}$ \cite{belle_1} is now well established \cite{fermilab_1}. \babar\ has seen evidence for the decay $X(3872)\rightarrow\jpsi\gamma$, which implies positive $C$-parity \cite{babar_radiative}. A variety of theoretical interpretations \cite{overview} exist for this state, including conventional charmonium interpretations \cite{barnes} and exotic QCD proposals such as a \bfDzDzb molecule \cite{molecule}. While \bfDzDzb molecular proposals can accommodate decays to $\jpsi\gamma$, the branching fraction for decays to $\psitwos\gamma$ is expected to be very small \cite{swanson}. These models allow for the possibility of an admixture of a \bfDzDzb bound state with, for example, a \ccbar meson. Because the $\chi_{c1}(2P)$ state potentially decays to $\psitwos\gamma$ at a rate many times higher than to $\jpsi\gamma$, the decay $X(3872)\rightarrow\psitwos\gamma$ could be enhanced due to \ccbar-\bfDzDzb mixing.

We present a study of the decay $B\rightarrow X K$, where the notation $X$ represents any state decaying radiatively to $\jpsi\gamma$ or $\psitwos\gamma$ (the $\chi_{c1,2}$ and $X(3872)$ states in particular), and $K$ encompasses \Kpm, \KS, $K^{*\pm}(892)$ and $K^{*0}(892)$. We consider \jpsi mesons decaying to $e^{+}e^{-}$ or $\mu^{+}\mu^{-}$, and \psitwos decaying to $e^{+}e^{-}$, $\mu^{+}\mu^{-}$ or $J/\psi\pi^{+}\pi^{-}$. Kaons are required to decay to final states consisting of charged particles; $\KS\rightarrow\pi^{+}\pi^{-}$, $\Kstarpm\rightarrow\KS(\pi^{+}\pi^{-})\pi^{\pm}$, and $\Kstarz\rightarrow\Kpm\pi^{\mp}$.

The data sample for this analysis consists of $(465\pm5)$ million \BB pairs collected with the \babar\ detector at the \pep2 asymmetric \epem collider at SLAC. This represents 424\invfb of data taken at the \FourS resonance. The \babar\ detector is described in detail elsewhere \cite{babar_nim}. The event selection, determined independently from the data, is based on Monte Carlo (MC) simulated events with the aim of maximizing significance.

The $\jpsi$ candidates are formed using pairs of leptons whose invariant mass is in the range (2.96,3.15) \gevcc for electrons (including bremsstrahlung photons) and (3.06,3.13) \gevcc for muons. For $\psitwos\rightarrow\ell^{+}\ell^{-}$, the candidate invariant masses are required to be in the range (3.61,3.73) \gevcc for electrons or (3.65,3.72) \gevcc for muons. The $\psitwos\rightarrow\jpsi\pi^{+}\pi^{-}$ candidates are composed of $\jpsi$ candidates decaying as described but with a tighter mass requirement of (3.01,3.15) \gevcc for the \epem decay mode. To form a $\psitwos$ candidate, the $\jpsi$ candidate is mass-constrained to the nominal PDG value \cite{pdg} and combined with a pair of oppositely charged tracks requiring (0.4,0.6) \gevcc and (3.68,3.69) \gevcc for the dipion and \psitwos invariant masses, respectively. All four final decay particles are constrained to the same decay vertex. Electrons are identified by a likelihood-based selector with $>92\%$ efficiency and negligible fake rate. Muons are selected by a neural net process with $>85\%$ efficiency and a $\pi$($K$) fake rate of $<6\%$ $(<10\%)$. Pions are drawn from the list of all charged tracks in the event.

We reconstruct $X\rightarrow\ccbar\gamma$ candidates from a mass-constrained \jpsi(\psitwos) candidate combined with a photon with an energy greater than $30(100)$ \mev. Additional selection criteria are applied to the shape of the lateral distribution ($0.001<LAT<0.5$) \cite{photon_lat} and azimuthal asymmetry (as measured by the Zernike moment $A_{42}<0.1$)\cite{photon_a42} of the photon-shower energy. For $X\rightarrow\jpsi\gamma$, the radiative $\gamma$ candidate is rejected if, when combined with any other $\gamma$ from the event, it has an invariant mass consistent with the $\pi^{0}$ mass, $124<m_{\gamma\gamma}<146$\mevcc.

The \KS candidates are required to be within $\pm17$\mevcc of the nominal \KS mass \cite{pdg}, and the significance of the distance of the reconstructed decay vertex from the primary vertex must be greater than $3.7$ standard deviations $(\sigma)$. The excited kaons are required to have an invariant mass within the range $0.7 < m(\Kstar) < 1.1$ \gevcc. For \KS, \Kstarpm, and \Kstarz candidates associated with $X\rightarrow\psi(2S)\gamma$, additional requirements are placed on the $\chi^{2}$ vertex probability of the kaon, $P(\chi^{2})>0.001, 0.02$ and $0.002$, respectively. Kaons are chosen by a likelihood-based selector with an efficiency of $\sim 95\%$ and misidentification rates of $\sim 5\%$, $\sim 4\%$, and $<10\%$ for $\pi$, $\mu$, and $p$ respectively, over the momentum range in this analysis.

We form the final \B candidate from an $X$ candidate and a kaon constrained to originate from the same vertex. To identify \B candidates, we use two kinematic variables, $m_{B}$ and $m_{\textnormal{miss}}$. The unconstrained mass of the reconstructed \B candidate is $m_{B} = \sqrt{E_{B}^{2}/c^{4}-p_{B}^{2}/c^{2}}$, where $E_{B}$ and $p_{B}$ are obtained by summing the energies and momenta of the particles in the candidate $B$ meson. The missing mass is defined as $m_{\textnormal{miss}}=\sqrt{(p_{\epem}-\hat{p}_{B})^{2}/c^{2}}$, where $p_{\epem}$ is the four-momentum of the beam \epem system and $\hat{p}_{B}$ is the four-momentum of the \B candidate after applying a \B mass constraint. For $X\rightarrow\jpsi(\psitwos)\gamma$ events, we require $m_{B}$ to be within $^{+30}_{-36}$ ($\pm20$)\mevcc of the nominal \B mass \cite{pdg}. Our \B candidate selection is further refined by imposing criteria on the $\chi^{2}$ probability for the \B vertex; for all $X\rightarrow\jpsi\gamma$ modes $P(\chi^{2})>0.0001$, and for $X\rightarrow\psitwos\gamma$ modes, $P(\chi^{2})>0.01$, $0.002$, and $0.05$ for the $K^{\pm}$, \KS, and \Kstar modes, respectively. The ratio of the second and zeroth Fox-Wolfram moments $(R_{2}<0.45)$ \cite{fox_wolfram} is used to separate isotropic \B events from continuum background events. Once a \B candidate has been established, it and its daughter decays are refit with the \B mass constrained to the known value \cite{pdg}.

We perform a one(two)-dimensional unbinned extended maximum-likelihood (UML) fit to $m_{\textnormal{miss}}$ (and $m_{K^{*}}$, if applicable), and use the $_{s}\textnormal{Plot}$ formalism \cite{splots} to project our signal events into $m_{X}$, the invariant mass of the $X$ candidate. This is a background-subtraction technique by weighting each event based on how signal- or background-like it is. The $_{s}\textnormal{Plot}$ displays the number of $B\rightarrow X K$ signal-like events as a function of $m_{X}$. We extract the signal yield for a given decay mode by fitting this resultant $m_{X}$ distribution with shapes for signal and background determined from MC simulation.

The signal event probability density functions (PDFs) are determined from MC-simulated $\B\rightarrow\chicone K$ and $\B\rightarrow X(3872) K$ events. Only reconstructed events exactly matching the generated decay chain particles are used to parameterize the signal PDFs. The PDF shapes for $\B\rightarrow\chi_{c2} K$ are the same as for $\chi_{c1}$, with the below-noted exception of the $m_{X}$ distribution. The $m_{\textnormal{miss}}$ distribution is modeled with a Crystal Ball function \cite{crystal_ball}, $m_{X}$ with a single Gaussian for the $\chi_{c2}$ decay modes and narrower core Gaussian plus a second wider Gaussian sharing the same mean for all other signal modes, and $m_{K^{*}}$ with the convolution of a Breit-Wigner and a Gaussian.

The background PDFs are determined from fits to generic \BpBm, \BzBzb, \qqbar, and \tautau MC samples, and are dominated by events from \BB decays that include a \jpsi or \psitwos in their decay chain. For the $\Bpm\rightarrow X \Kpm$ and $\Bz\rightarrow X \KS$ decay modes, the background in $m_{\textnormal{miss}}$ consists of two parts: a non-peaking combinatoric component modeled with an ARGUS function \cite{argus}, and a peaking component that shares the Crystal Ball parameterization used for signal events. These backgrounds are modeled as linear in $m_{X}$. The \Kstar decay modes have three background components: events that peak in $m_{\textnormal{miss}}$ but are flat in $m_{K^{*}}$ (``non-resonant'') and vice versa (``\Kstar combinatoric''), and those that do not peak in either distribution (``combinatoric''). The peaking $m_{\textnormal{miss}}$ and $m_{\Kstar}$ distributions use the same parameterization and values found by fitting to the signal MC sample. The non-peaking $m_{\textnormal{miss}}$ distributions are fit with an ARGUS function, while the non-peaking $m_{\Kstar}$ distribution is modeled with a linear function. Both combinatoric background types are flat in $m_{X}$, while the non-resonant backgrounds (typically $B\rightarrow X K\pi$) have a flat and peaking component in $m_{X}$. However, because none of these background events are signal-like in both $m_{\textnormal{miss}}$ and $m_{\Kstar}$, they are not present in the $_{s}\textnormal{Plot}$ projection in $m_{X}$.

To account for potential differences between data and MC, the values for the $m_{\textnormal{miss}}$ ARGUS and $m_{X}$ linear parameters for the background events are left as free parameters in the final fit to data. We also allow the height of the $m_{X}$ Gaussian peaks to float, which we use to derive the number of signal events.

The effectiveness of the signal extraction method is validated on fully-simulated MC events for $\chi_{c1,2}$ and $X(3872)$ signal events, with random samples generated from the MC background distribution. Successful performance of the fit is verified on simulated datasets assuming the number of signal and background events from the known branching fractions and efficiencies. We apply small corrections ($<5\%$) to account for bias in the results of the MC fit validation.

We determine the efficiency from the fraction of the events generated in MC simulation that survive the analysis selection criteria and are returned by the fitting procedure. We calculate the branching fraction for each decay mode using $\BR(\B\rightarrow X K)= N_{S}/(N_{\BB}\times\epsilon\times f)$ where $N_{S}$ is the bias-corrected number of signal events from the fit to the $m_{X}$ $_{s}\textnormal{Plot}$, $N_{\BB}$ is the number of $\BB$ pairs in the data set, $\epsilon$ is the total signal extraction efficiency, and $f$ represents all secondary branching fractions. The fit results, efficiencies, and derived branching fractions are summarized in Table \ref{tab_results}.

\begin{table}[htb]
\hspace{-1.0cm}
\caption {Summary of the analysis results. $N_{S}$ is the bias-corrected number of signal events extracted from the $m_{X}$ $_{s}\textnormal{Plot}$, $\sigma$ is the total significance of the signal yield $N_{S}$ measured in standard deviations (statistical and systematic uncertainties combined in quadrature) from the null result, $\epsilon$ is the total efficiency for the decay mode, and derived $\BR$ is the measurement (with 90\% confidence level upper limit \cite{90ul}) of $\BR(B\rightarrow\chi_{c1,2} K)$ or $\BR(B\rightarrow X(3872) K)\cdot\BR(X\rightarrow\ccbar\gamma)$. Uncertainties are statistical and systematic, respectively.}
\begin{tabular}{l|c|c|c|c}
\hline \hline
Decay & $N_{S}$ & $\sigma$ & $\epsilon(\%)$ & Derived $\BR$\\
\hline\noalign{\vskip1pt}
\multicolumn{4}{l}{$\chi_{c1}$} & $\times10^{-4}$\\
\hline
$\chi_{c1} \Kpm$ & $1018\pm34\pm14$ & $28$ & $11.0$ & $4.5\pm0.1\pm0.3$\\
$\chi_{c1} K^{0}$ & $242\pm16\pm5$ & $14$ & $8.7$ & $4.2\pm0.3\pm0.3$\\
$\chi_{c1} \Kstarpm$ & $71\pm13\pm8$ & $4.7$ & $5.7$ & $2.6\pm0.5\pm0.4$\\
$\chi_{c1} \Kstarz$ & $255\pm25\pm11$ & $9.5$ & $7.9$ & $2.5\pm0.2\pm0.2$\\
\hline\noalign{\vskip1pt}
\multicolumn{4}{l}{$\chi_{c2}$} & $\times10^{-5}$\\
\hline
$\chi_{c2} \Kpm$ & $14.0\pm7.9\pm1.1$ & $1.8$ & $12.3$ & $1.0\pm0.6\pm0.1(<1.8)$\\
$\chi_{c2} K^{0}$ & $6.1\pm3.9\pm1.1$ & $1.5$ & $11.1$ & $1.5\pm0.9\pm0.3(<2.8)$\\
$\chi_{c2} \Kstarpm$ & $1.2\pm4.7\pm6.1$ & $0.2$ & $4.2$ & $1.1\pm4.3\pm5.5(<12)$\\
$\chi_{c2} \Kstarz$ & $38.8\pm10.5\pm1.1$ & $3.7$ & $8.3$ & $6.6\pm1.8\pm0.5$\\
\hline\noalign{\vskip1pt}
\multicolumn{4}{l}{$X(3872)(\jpsi\gamma)$} & $\times10^{-6}$\\
\hline
$X \Kpm$ & $23.0\pm6.4\pm0.6$ & $3.6$ & $14.5$ & $2.8\pm0.8\pm0.1$\\
$X K^{0}$ & $5.3\pm3.6\pm0.2$ & $1.5$ & $11.0$ & $2.6\pm1.8\pm0.2(<4.9)$\\
$X \Kstarpm$ & $0.6\pm2.3\pm0.1$ & $0.3$ & $6.9$ & $0.7\pm2.6\pm0.1(<4.8)$\\
$X \Kstarz$ & $2.8\pm5.2\pm0.4$ & $0.5$ & $10.4$ & $0.7\pm1.4\pm0.1(<2.8)$\\
\hline\noalign{\vskip1pt}
\multicolumn{4}{l}{$X(3872)(\psitwos\gamma)$} & $\times10^{-6}$\\
\hline
$X \Kpm$ & $25.4\pm7.3\pm0.7$ & $3.5$ & $10.4$ & $9.5\pm2.7\pm0.6$\\
$X K^{0}$ & $8.0\pm3.9\pm0.5$ & $2.0$ & $8.4$ & $11.4\pm5.5\pm1.0(<19)$\\
$X \Kstarpm$ & $1.9\pm2.9\pm2.9$ & $0.5$ & $5.0$ & $6.4\pm9.8\pm9.6(<28)$\\
$X \Kstarz$ & $-1.4\pm3.3\pm0.3$ & $-$ & $6.7$ & $-1.3\pm3.1\pm0.3(<4.4)$\\
\hline
\end{tabular}
\label{tab_results}
\end{table}

For most of the $B\rightarrow X(3872) K$ decay channels, the largest source of systematic uncertainty affecting the signal yield comes from the uncertainty in the true $X(3872)$ mass and width ($\sim$2$\%$ for the \Kpm modes). In the case of the \Kpm and \KS decay modes for $X(3872)\rightarrow\psitwos\gamma$, an alternate parametrization of the $m_{X}$ shape was considered for background events, as indicated by the MC simulation. A correction equal to half the difference between the results of the two background model choices, with a systematic error equal to this amount, is applied to the final result. This is the largest yield-related systematic uncertainty for the $X(3872)(\psitwos\gamma)\Kpm$ mode ($\sim$2$\%$). For $B\rightarrow\chi_{c1,2} K$, uncertainty in the fit bias, PDF parameters, and MC/data differences for the mean value $m_{X}$ for signal events all contribute in varying though roughly equal amounts.

Regarding systematic uncertainties related to the branching fraction calculations, one of the main contributors is the total uncertainty associated with the identification of all particle types $(\sim$4$\%)$. The uncertainty in secondary branching fractions, beyond the control of this analysis, is the dominant systematic uncertainty for $\BR(B\rightarrow\chi_{c1} K)$ and $\BR(B^{0}\rightarrow\chi_{c2} K^{*0})$ $(\sim$6$\%)$. Effects from tracking, photon corrections and $B$ counting are also considered, but are all less than $2\%$.

\begin{figure}
\begin{center}
\epsfig{file={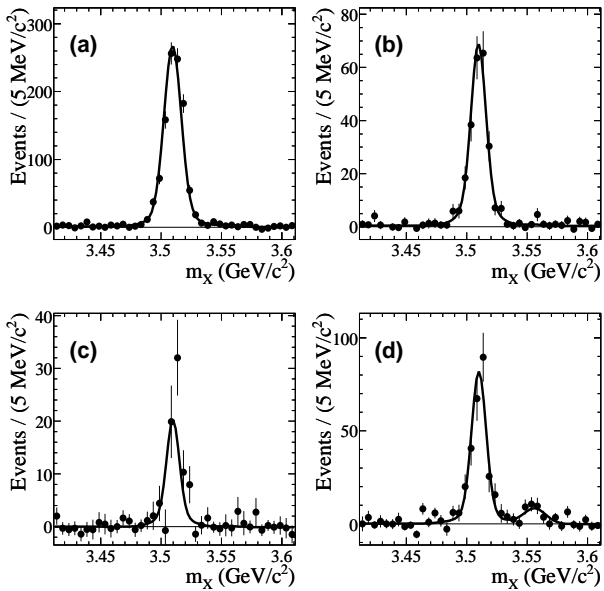},width={3.2in},height={3.2in}}
\caption{$_{s}\textnormal{Plot}$ of the number of signal events versus $m_{X}$ for (a) $\Bpm\rightarrow\chi_{c1,2}\Kpm$, (b) $\Bz\rightarrow\chi_{c1,2}\KS$, (c) $\Bpm\rightarrow\chi_{c1,2}\Kstarpm$, and (d) $\Bz\rightarrow\chi_{c1,2}\Kstarz$. The solid curve is the fit to the data.}
\label{fig_chic}
\end{center}
\end{figure}
\begin{figure}
\begin{center}
\epsfig{file={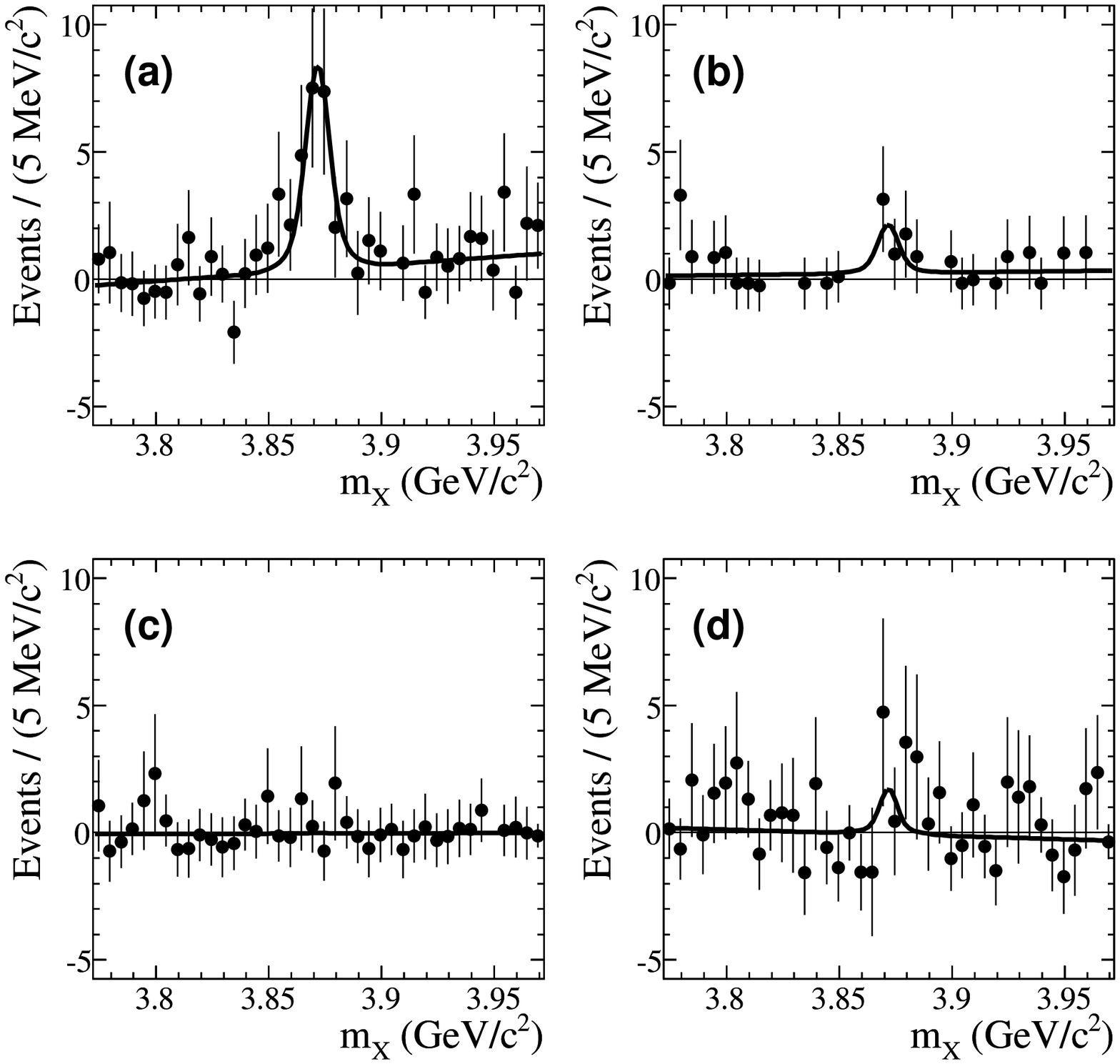},width={3.2in},height={3.2in}}
\caption{$_{s}\textnormal{Plot}$ of the number of extracted signal events versus $m_{X}$ for (a) $\Bpm\rightarrow X(3872)\Kpm$, (b) $\Bz\rightarrow X(3872)\KS$, (c) $\Bpm\rightarrow X(3872)\Kstarpm$, and (d) $\Bz\rightarrow X(3872)\Kstarz$, where $X(3872)\rightarrow\jpsi\gamma$. The solid curve is the fit to the data.}
\label{fig_x3872_jpsi}
\end{center}
\end{figure}
\begin{figure}
\begin{center}
\epsfig{file={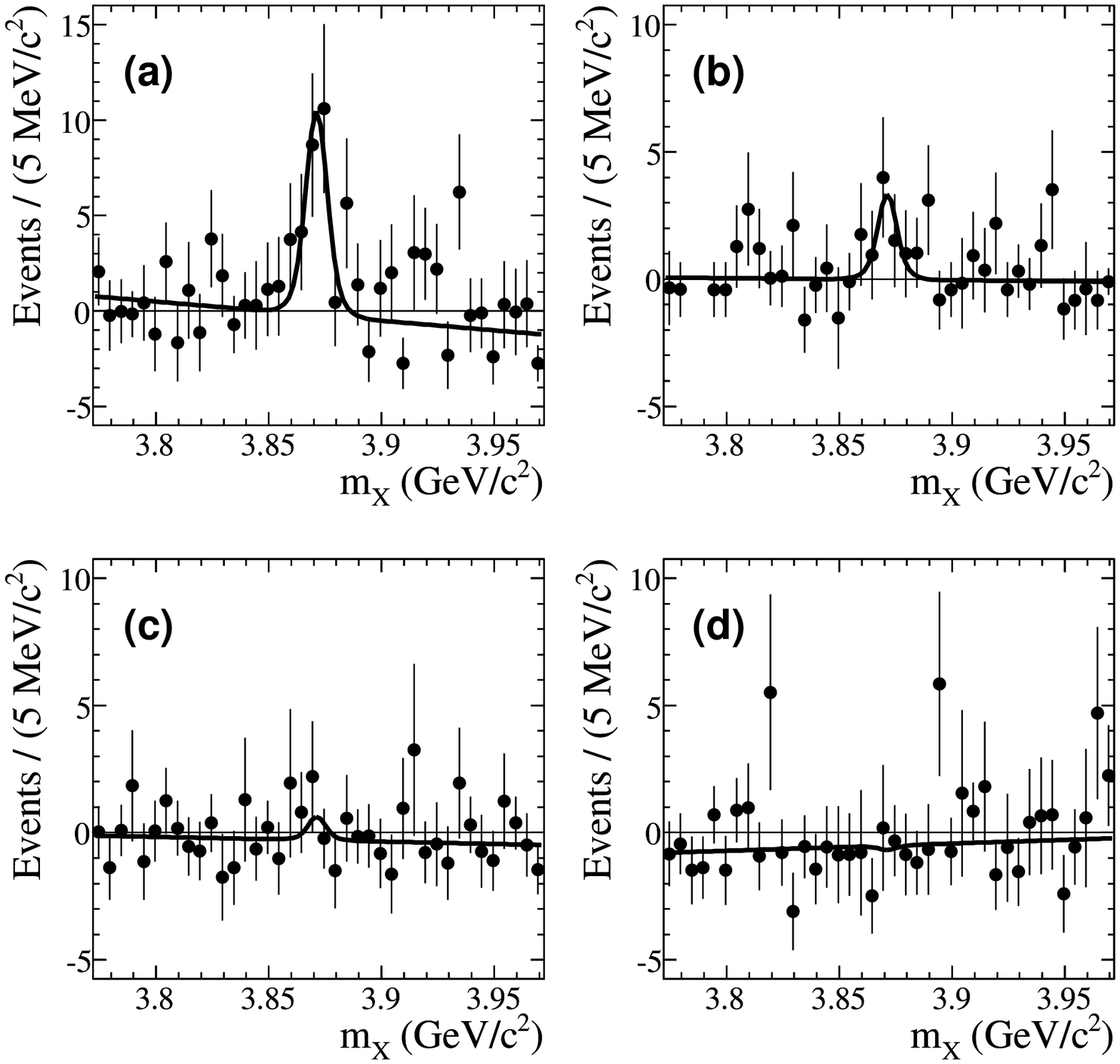},width={3.2in},height={3.2in}}
\caption{$_{s}\textnormal{Plot}$ of the number of extracted signal events versus $m_{X}$ for (a) $\Bpm\rightarrow X(3872)\Kpm$, (b) $\Bz\rightarrow X(3872)\KS$, (c) $\Bpm\rightarrow X(3872)\Kstarpm$, and (d) $\Bz\rightarrow X(3872)\Kstarz$, where $X(3872)\rightarrow\psitwos\gamma$. The solid curve is the fit to the data.}
\label{fig_x3872_psi2s}
\end{center}
\end{figure}

Figure \ref{fig_chic} shows the fit to $m_{X}$ in the mass range $3.411<m_{X}<3.611$ \gevcc. We observe all of the expected $B\rightarrow\chi_{c1} K$ decay modes, in good agreement with previous measurements. We find $3.7\sigma$ evidence for $\Bz\rightarrow\chi_{c2}\Kstarz$, and set upper limits for the remaining $B\rightarrow\chi_{c2} K$ decays.

Fits to $m_{X}$ in the range $3.772 < m_{X} < 3.972$ \gevcc are shown in Fig. \ref{fig_x3872_jpsi} for decays to $\jpsi\gamma$. We confirm evidence for the decay $X(3872)\rightarrow\jpsi\gamma$ in $\Bpm\rightarrow X(3872)\Kpm$, measuring $\BR(\Bpm\rightarrow X(3872) \Kpm)\cdot\BR(X(3872)\rightarrow\jpsi\gamma) = (2.8 \pm 0.8 (stat.) \pm 0.1 (syst.))\times10^{-6}$ with a significance of $3.6\sigma$. This value is in good agreement with the previous \babar\ result \cite{babar_radiative}, which it supersedes, and represents the most precise measurement of this branching fraction to date. We find no significant signal in the other decay modes.

Figure \ref{fig_x3872_psi2s} shows the fit to the $m_{X}$ distribution for $X\rightarrow\psitwos\gamma$ in the range $3.772 < m_{X} < 3.972$ \gevcc. In our search for $X(3872)\rightarrow\psitwos\gamma$ in $\Bpm\rightarrow X(3872)\Kpm$, we find the first evidence for this decay with a significance of $3.5\sigma$. We derive $\BR(\Bpm\rightarrow X(3872) \Kpm)\cdot\BR(X(3872)\rightarrow\psitwos\gamma) = (9.5 \pm 2.7 (stat.) \pm 0.6 (syst.))\times10^{-6}$. We find no significant signals in the other decay modes.

To search for other new resonances, the $m_{X}$ invariant mass window is extended up to the kinematic limit. Even after combining all decay modes together, there are no indications of any further signals above the prominent $X(3872)$ peak.

In summary, we present first evidence for the decay $X(3872)\rightarrow\psi(2S)\gamma$, and updated measurements of the $X(3872)\rightarrow\jpsi\gamma$ and $B\rightarrow\chi_{c1,2} K$ decays. We find evidence for the factorization-suppressed \cite{colangelo} decay $\Bz\rightarrow\chi_{c2}\Kstarz$, but see no evidence for other $\chi_{c2} K$ decays. Taking the statistical and systematic errors in quadrature, we find a ratio of $\frac{\BR(X(3872)\rightarrow\psitwos\gamma)}{\BR(X(3872)\rightarrow\jpsi\gamma)} = 3.4\pm1.4$. This relatively large branching fraction for $X(3872)\rightarrow\psitwos\gamma$ is generally inconsistent with a purely \bfDzDzb molecular interpretation of the $X(3872)$, and possibly indicates mixing with a significant \ccbar component.

We are grateful for the excellent luminosity and machine conditions
provided by our \pep2\ colleagues, 
and for the substantial dedicated effort from
the computing organizations that support \babar.
The collaborating institutions wish to thank 
SLAC for its support and kind hospitality. 
This work is supported by
DOE
and NSF (USA),
NSERC (Canada),
CEA and
CNRS-IN2P3
(France),
BMBF and DFG
(Germany),
INFN (Italy),
FOM (The Netherlands),
NFR (Norway),
MES (Russia),
MEC (Spain), and
STFC (United Kingdom). 
Individuals have received support from the
Marie Curie EIF (European Union) and
the A.~P.~Sloan Foundation.


\begin{thebibliography}{99}
\bibitem{90ul}
The upper limit (UL) is calculated from $\int^{UL}_{0}G(x)/\int^{+\infty}_{0}G(x)=0.9$, where $G(x)$ is a Gaussian with mean equal to the central value of the branching fraction measurement and standard deviation equal to the total uncertainty.
\end{thebibliography}
\end{document}